\documentclass[twocolumn,amsmath,amssymb]{revtex4}

\usepackage{graphicx}
\usepackage{graphics}
\usepackage{subfigure}
\usepackage{dcolumn}
\usepackage{bm}
\usepackage{txfonts}

\begin{document}

\title{Polarization, Abstention, and the Median Voter Theorem}
\author{Matthew I. Jones$^{1}$}
\email{Matthew.I.Jones.GR@dartmouth.edu}
\author{Antonio D. Sirianni$^{2}$}
\email{Antonio.D.Sirianni@dartmouth.edu}
\author{Feng Fu$^{1,3}$}
\email{fufeng@gmail.com}

\affiliation{ $^1$Department of Mathematics, Dartmouth College, Hanover, NH 03755, USA\\
$^2$ Department of Sociology, Dartmouth College, Hanover, NH 03755, USA\\
$^3$Department of Biomedical Data Science, Geisel School of Medicine at Dartmouth, Lebanon, NH 03756, USA}

\date{\today}

\begin{abstract}
The median voter theorem has long been the default model of voter behavior and candidate choice.  While contemporary work on the distribution of political opinion has emphasized polarization and an increasing gap between the ``left" and the ``right" in democracies, the median voter theorem presents a model of anti-polarization: competing candidates move to the center of the ideological distribution to maximize vote share, regardless of the underlying ideological distribution of voters. These anti-polar results, however, largely depend on the ``singled-peakedness" of voter preferences, an assumption that is rapidly loosing relevance in the age of polarization. This article presents a model of voter choice that examines three potential mechanisms that can undermine this finding: a relative cost of voting that deters voters who are sufficiently indifferent to both candidates, ideologically motivated third-party alternatives that attract extreme voters, and a bimodal distribution of voter ideology. Under reasonable sets of conditions and empirically observed voter opinion distributions, these mechanisms can be sufficient to cause strategically-minded candidates to fail to converge to the center, or to even become more polarized than their electorate.
\end{abstract}

 \maketitle
\section{Introduction}

When is it rational for two strategically-motivated candidates to deviate from the ideological center in a general election? Spatial models of economic competition have long served as a baseline model for political agendas and electoral outcomes \cite{Hotelling1929,Downs1957}. In their simplest form, every voter's political preference is captured along a one-dimensional space, and each voter chooses the candidate (typically out of two) who is most proximate to them in the one-dimensional ideological space. Accordingly, each candidate rationally selects a point which maximizes their share of votes. The main result is well-known, two competing and self-interested candidates are at equilibrium when their political positions are equal to the opinion of the median voter. 

A simple one-dimensional, two-candidate model of elections ignores many possible complications that have been addressed by political scientists since the conception of the median voter theorem. There may be more than two candidates, or a third option may enter depending on the ideological alignment of the two main candidates \cite{Palfrey1984}. The ideological space that candidates are competing on may be multi-dimensional \cite{Davis1966,Hinich1970}. Voters may also have probabilistic rather than deterministic voting rules, which can shift the point of candidate ideological convergence from the median to the center \cite{Banks2005,McKelvey2006}. Candidates may not be purely concerned with winning, and gain more utility from winning with a specific ideological position \cite{Kollman1992}. 

We focus on a set of three main complications that are undeniably present in the United States, but have not yet been examined in tandem. First, we consider the influence of ideologically motivated third-party candidates. While third-party voting has been on the decline in the United States \cite{Hirano2007}, voting for non-competitive third-party candidates still occurs as an expression of cynicism or distrust with the the larger political system \cite{Peterson1998}, often types at levels that sway the results of major elections \cite{Allen2005}. 

Second, elections in the United States of America typically feature large numbers of eligible voters who stay at home \cite{Franklin2004}. The reasons for voter abstention have been well-studied. Voters may choose to abstain in protest if they feel that both candidates are unacceptably far away from their preferences, or if they are indifferent between candidates \cite{Hinich1969,Enelow1984,Anderson1992,Adams2006,Thurner2000}. The costs of voting may influence turn out: averse weather conditions lowers voting rates \cite{Gomez2007}, while same-day voter registration decreases costs and increases turnout \cite{Fenster1994}.

These first two sources of voter-abstention have been empirically examined using data from U.S. elections  \cite{Plane2004,Adams2006,Poole1984}. Their combination leads to the perpetually discussed dilemma of winning over the center or appealing to the `base' when determining what candidates or platforms to field for a general election \cite{Abramson1992}. On one hand extreme candidates might cede the center to the opponent (in line with the assumptions of the median voter theorem), and on the other hand extremist voters may behave irrationally and stay at home rather than casting a vote for the candidate who is closest to them ideologically. An additional concern is that extremist candidates, while energizing their own base, may increase turnout for people who are extremely opposed to their agenda as well \cite{Hall2018}.

The final mechanism we consider in our model is polarization. Polarization has been examined extensively by political scientists \cite{Fiorina2008}, sociologists \cite{Baldassarri2007}, and economists \cite{Dixit2007}, and its empirical scope and potential causes have been the focus of impressive studies by information scientists \cite{Conover2011}, and computational social scientists \cite{Bail2018}, but it's implications for rational choice voting models and candidate competition are rarely considered \cite{Grosser2014,Wang2020}. Moreover, median voter-type results often are predicated on the single-peakedness of voter preferences \cite{Black1958}, which is a problematic assumption during times of polarization. The median voter model can be seen as a ``bottom-up" process that brings the political preferences of rational candidates in line with the more centrist preferences of the electorate. It is a model of anti-polarization \cite{Grosser2014}, but its limitations have become apparent in the current political climate.

Given these three variables: voter tendencies towards third-party candidates, staying home, and polarized beliefs, we are primarily interested in whether specific combinations will motivate strategic candidates to pursue divergent ideological strategies. Given the growing polarization in the U.S. Electorate \cite{Webster2017}, it is important to consider the conditions necessary for candidates to follow voters in their drift to extreme positions in the short term.

Our approach builds on more parsimonious models of voter choice by allowing voters to either choose one of the two main strategically-motivated candidates, an ideologically motivated third-party candidate, or stay home altogether. We also consider the ideological distribution of the voter electorate as a proxy for political polarization. Following earlier advances in the voter choice literature, our approach treats voting as a stochastic rather than deterministic process \cite{Coughlin1992}: the odds of a voter choosing a candidate increase with their relative ideological proximity, but it is never a certainty. This analytical decision is thought to better model voter uncertainty \cite{Burden1997}. A stochastic voting model has shown that preferential skew does lead to non-median outcomes \cite{Coughlin1984,Comanor1976,Hinich1976}, but these models still have one unique equilibrium.

In our analysis, we systematically vary the ideological distribution of voters, the appeal of ideologically-motivated third party candidates on the far ends of the political spectrum, and the appeal of staying home all together. We then map the conditions under which rational political candidates to fail to converge on the median ideological position, and also when candidates become more extreme than the electorate itself. We then analyze these dynamics with two empirically observed voter opinion distributions from the contemporary United States.

\section{Methods and Model}

\subsection{A Model of Voter Selection and Population Polarization}

Our model examines how a polarized population can influence the political positions of two strategically motivated candidates, who are purely interested in maximizing vote share. Building and integrating the aforementioned models of voter choice, we allow for the possibility that voter may either select a ideologically motivated and extreme candidate instead of a major-party candidate, or that voters may vote for neither candidate if they find their choices unappealing.

Our model considers how both the ideological distribution of the voters and voter tendencies to select one of the two major candidates should influence the political positioning of the two main candidates. These patterns change even when the median and mean voter position is fixed at the center of the distribution. Before discussing the results of our approach, we first outline the two main variable parts of the model: the distribution of the voters and the function that is used to map voter ideology to voter choice and behavior.

\subsection{Ideological Distribution of Voters}
We assume a single-dimensional ideological distribution of voters, $x$, on a scale from 0 (left) to 1 (right). We assume that voters made up of two sub-populations, consolidating around two ``peaks'' that are equidistant from the ideological center (0.5). The distance between the peaks is determined by $\alpha$, and the variance in position around these two peaks is determined by $\sigma^2$.

Mathematically this provides a population probability density function that is the sum of two normal distributions, $f(x)$:
\begin{equation}\label{eq:pdfparams}
    f(x) = c\Big[ \mathcal{N}(\frac{1}{2}+\alpha/2,\sigma^2) + \mathcal{N}(\frac{1}{2}-\alpha/2,\sigma^2) \Big]
\end{equation}
Where $\mathcal{N}(a,b^2)$ is the normal distribution with mean $a$ and variance $b^2$, and $c$ is a normalizing constant to ensure that $\int_{0}^1 f(x)dx = 1$.  This population is symmetric, and the median voter is always located at 0.5. Figure 1 illustrates this distribution. While we focus on our model on a hypothetical case where there are two balanced left-leaning and right-leaning subpopulations, the underlying ideological distribution of an actual population, which is not necessarily symmetric, can be calibrated using real voter data from any population of interest \cite{brown2021partisan}.


\begin{figure}
    \centering
    \includegraphics[width = 0.5\textwidth]{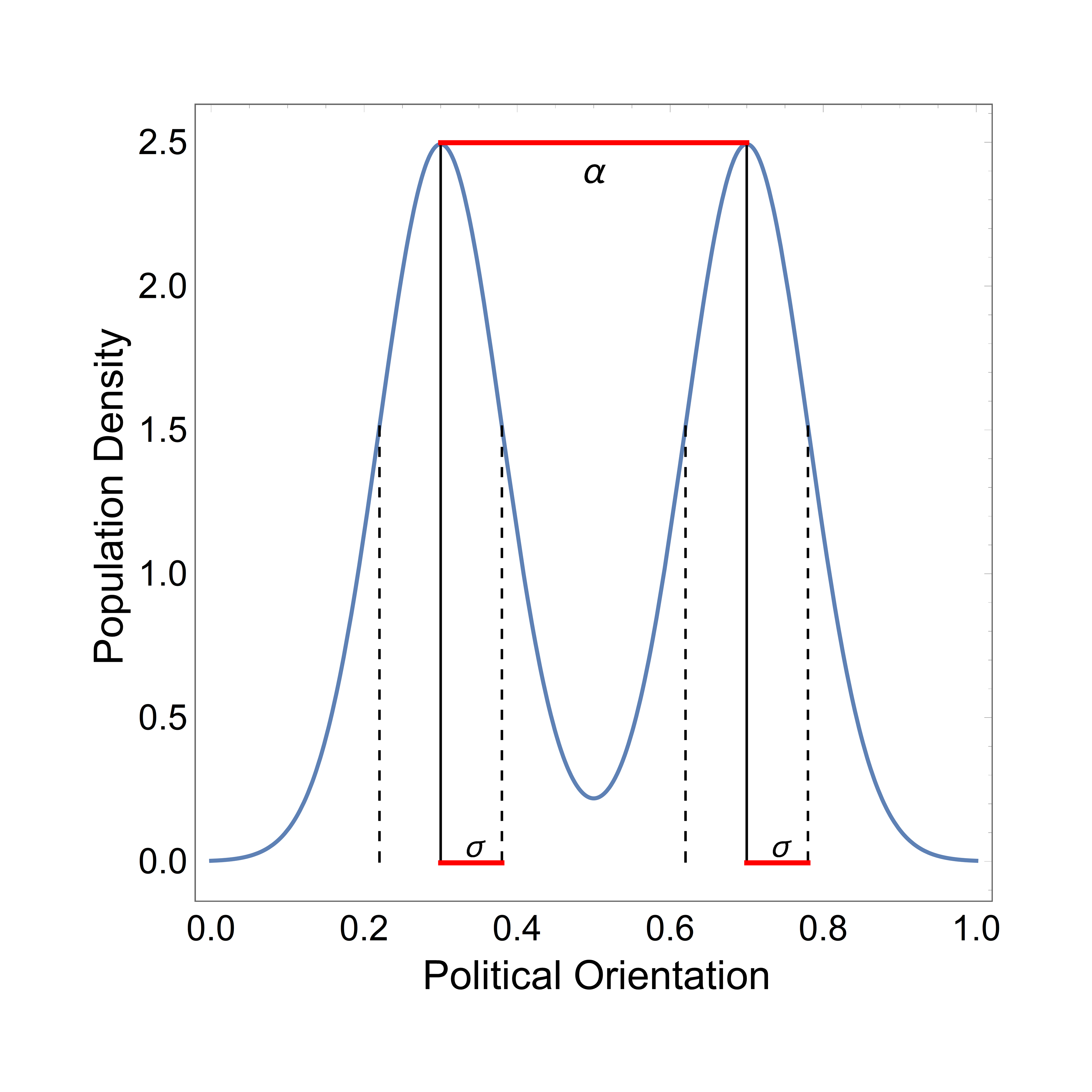}
    \caption{Ideological distribution of voters as a function of the two population parameters, $\alpha$ and $\sigma$. $\alpha$ is the distance between the two subpopulation centers, and $\sigma$ is the variance around these subpopulation centers. As $\sigma$ increases, the population distributions will become less pronounced and more diffuse.}
    \label{fig:PopParams}
\end{figure}

\subsection{Voter Choice Function}

When voters always select the most ideologically proximate candidate, both parties would still converge to the opinion of the median voter, which is fixed at 0.5. Varying the distribution of voters would have no effect on the strategic ideological positions of the candidates.

Yet in reality political candidates frequently express concern about losing their `base' when trying to appeal to the `center'. The threat of losing the base only occurs when voters have the option of either abstaining or selecting a third-party candidate that adapts a position in accordance with their ideology as opposed to vote maximization. Conversely, voters in the center may abstain if both candidates assume positions that are too extreme for them.



In our model, there are three variables that control voting behavior: pragmatism ($P$), which can be thought of as the appeal of voting for a two-party candidate, relative cost of voting ($Q$) which adjusts the voter tendency for staying home, and rebelliousness ($R$), which determines the appeal of third-party candidates. $P$ and $R$ are similar, and balance the candidate's preferences towards an ideologically motivated third-party selection or a more practical two-party selection. When voters are more ideologically equidistant from candidates, they should be more likely to stay home altogether. $Q$ is a multiplier for this, such that the utility a voter gets from not voting is a product of $Q$ and a voters' ideological indifference between the two candidates.


The behavior of the voter is determined by behavioral utilities calculated from the three above parameters, the ideological position of both of the major parties, and the ideological position of the voter in question.  

For an individual at $v$ and major candidates at $b$ and $r \in [0,1]$, we get the following utilities:
\begin{equation}
    \textrm{Vote Blue Utility} = u_B(b,v) = \frac{1}{|b-v|^P}
\end{equation}
\begin{equation}
    \textrm{Vote Red Utility} = u_R(r,v) = \frac{1}{|r-v|^P}
\end{equation}
\begin{equation}
    \textrm{Abstention Utility} = u_A(b,r,v) = (1-|(|b-v|-|r-v|)|)Q
\end{equation}
\begin{equation}
    \textrm{Vote Third Party Utility} = u_T(v) = \frac{1}{(1-v)^R}+\frac{1}{v^R}
\end{equation}

Each voter chooses from one of the four possible behaviors (vote for red, vote for blue, vote for third party, and abstain) with a probability that is proportional to each of their respective utilities. Figures 2 and 3 provide visual depictions of how voter behavior varies in the model as functions of voter and candidate ideology, respectively.

\section{Results}

\subsection{Voter Choice Dynamics}

Figure 2 shows voter utilities and corresponding probabilities for a set of parameters. The ``candidates'' have ideological positions of 0.3 and 0.7, somewhere between being completely polarized and converging to the middle, which roughly reflects two-party elections in the contemporary United States. To illustrate the model, we select a set of parameters for the proposed voter utility functions that lead to an intuitively plausible relationship between voter ideology and voter behavior. The values $P =2$, $Q=30$, and $R=1$ cause more ``extreme'' voters with an ideology closer to 0 or 1 to be more likely to select a third party candidate or stay home. Furthermore, the voters in the ideological valley between the two candidates are more likely to stay home, as they do not gain much of a relative benefit from either candidate.

Figure 3 also uses this set of ``common sense" decision parameters, but instead focused on the decision behavior of a single voter at a fixed ideological point, and examines how voter behavior corresponds to the ideological positions of the two main candidates. For a voter with an ideology of 0.5, a ``median voter", they become more likely to choose a blue or red candidate when one of them adopts a platform that is ideologically moderate.  They become more likely to abstain when both candidates choose more extreme candidate positions on either the same or opposing sides of the political spectrum.

\begin{figure*}[t]
    \centering
    \includegraphics[width=\textwidth]{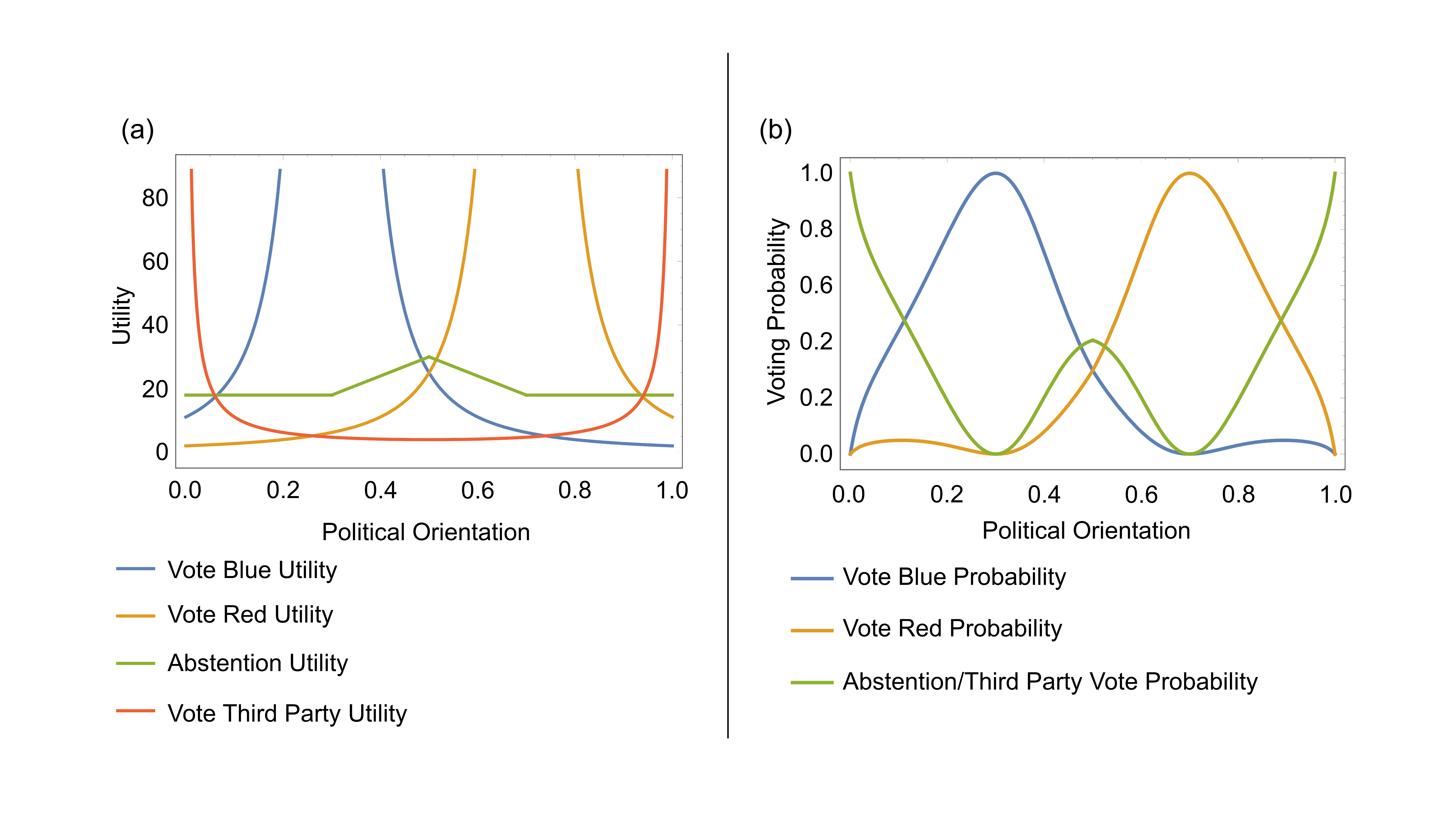}
    \caption{Voter utility and voting behavior of individuals across the entire political spectrum with two fixed political candidates. (a) shows the utility that a voter receives from different actions as a function of their position on the political spectrum, assuming candidate positions of 0.3 and 0.7 and a specific set of model parameters ($P=2$, $Q=30$, $R=1$). (b) maps these utilities into one of three behaviors: voting for the ``blue'' (left-leaning) candidate, the ``red'' (right-leaning) candidate, or voting for neither (staying home or selecting an ideologically motivated third-party candidate).}
    \label{fig:populationVoting}
\end{figure*}


\begin{figure*}[t]
    \centering
    \includegraphics[width=\textwidth]{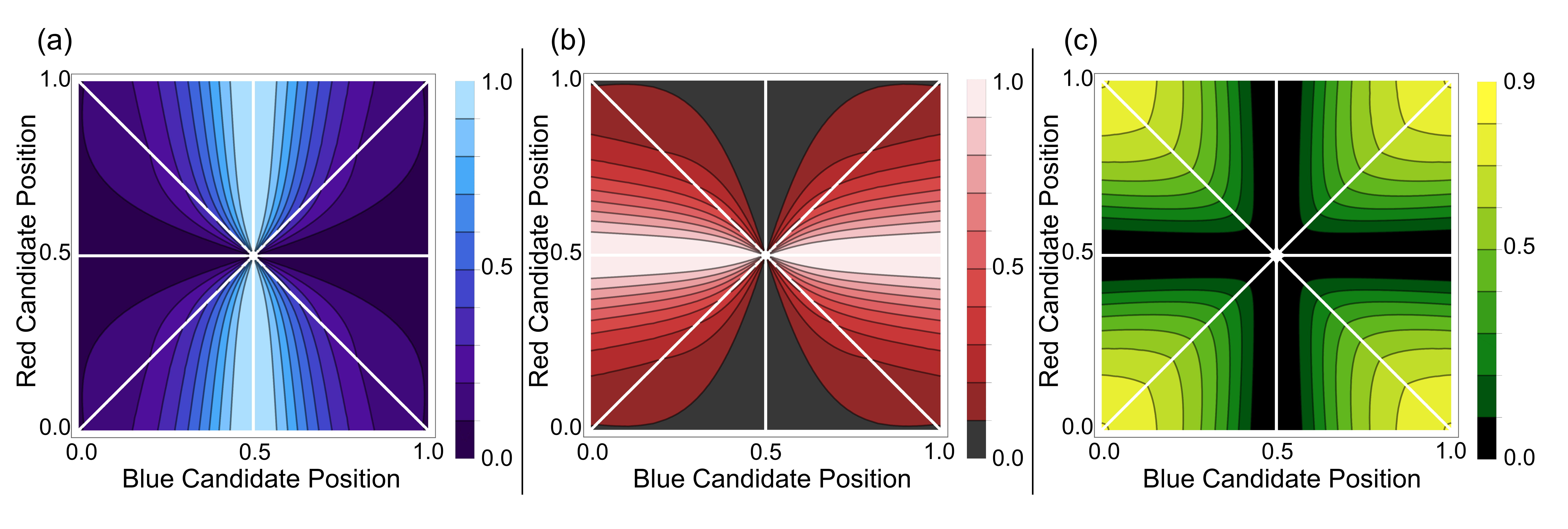}
    \caption{Varying the candidates positions influences a fixed voter's behavior. Each panel shows how the likelihood of a given voter behavior (voting `blue' in (a), voting `red' in (b), or voting for neither in (c)) changes as a function of the two political candidates stated ideological position, $[0, 1]^2$, assuming a voter ideology of $0.5$, a specific set of model parameters ($P=2, Q=30, R=1$).}
    \label{fig:individualVoting}
\end{figure*}

\subsection{Candidate Optimization}
Given that the behavior of each voter is stochastically determined as a function of their ideological position, the positions of the candidates, and the parameters of our model, we can determine the optimal ideological positions for two competing candidates who are motivated by maximizing vote share.
For an ideological space that stretches from 0 (on the left) to 1 (on the right), the liberal and conservative candidates are each seeking an ideological position (`$b$' or `$r$' respectively) that maximizes the value of one of the following integrals:

\begin{widetext}
\begin{equation}
    \textrm{Blue Votes} = v_B(b,r) =  \int_0^1 f(v) \frac{u_B(b,v)}{u_B(b,v)+u_R(r,v)+u_A(b,r,v)+u_T(v)} dv
\end{equation}
\begin{equation}
    \textrm{Red Votes} = v_R(b,r) = \int_0^1 f(v) \frac{u_R(r,v)}{u_B(b,v)+u_R(r,v)+u_A(b,r,v)+u_T(v)} dv
\end{equation}
\end{widetext}

The two major candidates' fictitious optimization process in response to voters' behavior can be described by the so-called adaptive dynamics~\cite{hofbauer1990adaptive,yang2020us}:

\begin{equation}
\begin{split}
\frac{db}{dt} & = \frac{\partial v_B(b,r)}{\partial b},\\
\frac{dr}{dt} & = \frac{\partial v_R(b,r)}{\partial r}.
\end{split}
\end{equation}
 
\begin{figure*}[t]
    \centering
    \includegraphics[width = \textwidth]{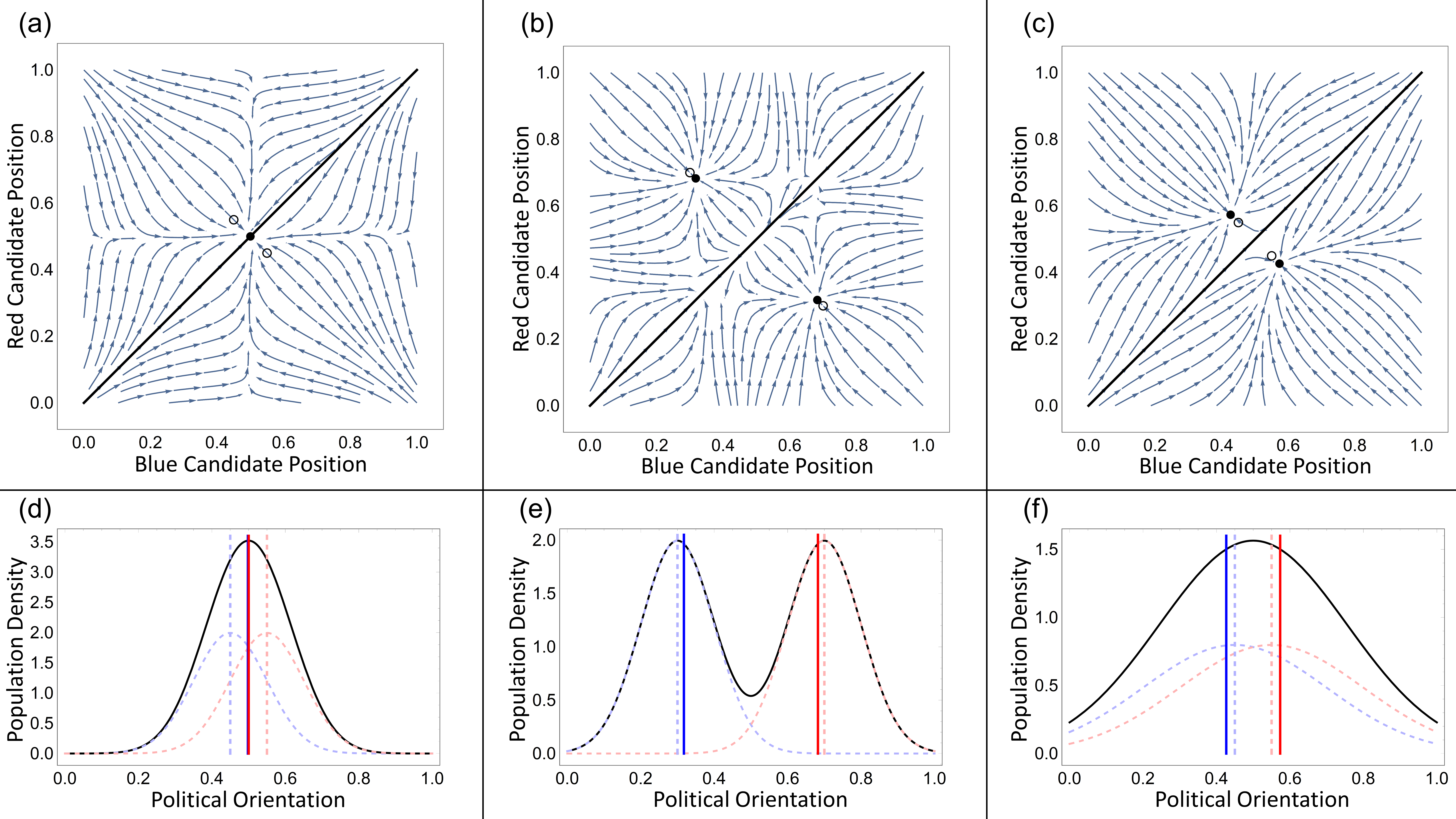}
    \caption{Optimal position of two competing major candidates (a)-(c) show how candidates will shift their position in a stream plot, with the black dot representing the candidates' equilibria, and the circles showing the subpopulation peaks. (d)-(e) show the corresponding populations in black, the two subpopulations with dashed curves, the subpopulation centers represented by dashed vertical lines, and the candidate equilibrium positions represented by solid vertical lines.
    We see three types of behavior: candidates converging to the median voter (a,d), candidates less polarized than the population (b,e), and candidates more polarized than the population (c,f).
    All plots use a reasonable set of parameters $P=2$, $Q=30$, and $R=1$.}
    \label{fig:samplemodel}
\end{figure*}

When voters choose the most ideologically proximate of the two competing candidates, both positions converge on the ideology of the median voter. Our model shows how this result does not necessarily hold when voters might choose to abstain or select a third party. In particular this can occur when the distribution of voter preferences is sufficiently bimodal. Figure \ref{fig:samplemodel} shows three different sample voter ideological distributions (d-e), and how two political candidates will adjust their ideological platform under a reasonable set of voter choice parameters for each (a-c).

With these three populations, candidate behavior varies from appealing to the median voter when competition is fierce in the high-density middle to being more polarized than the population as candidates work to protect their most extreme voters from a third party challenge.

\subsection{Candidate Positions and Voter Distributions}

\begin{figure*}[t]
    \centering
    \includegraphics[width =\textwidth]{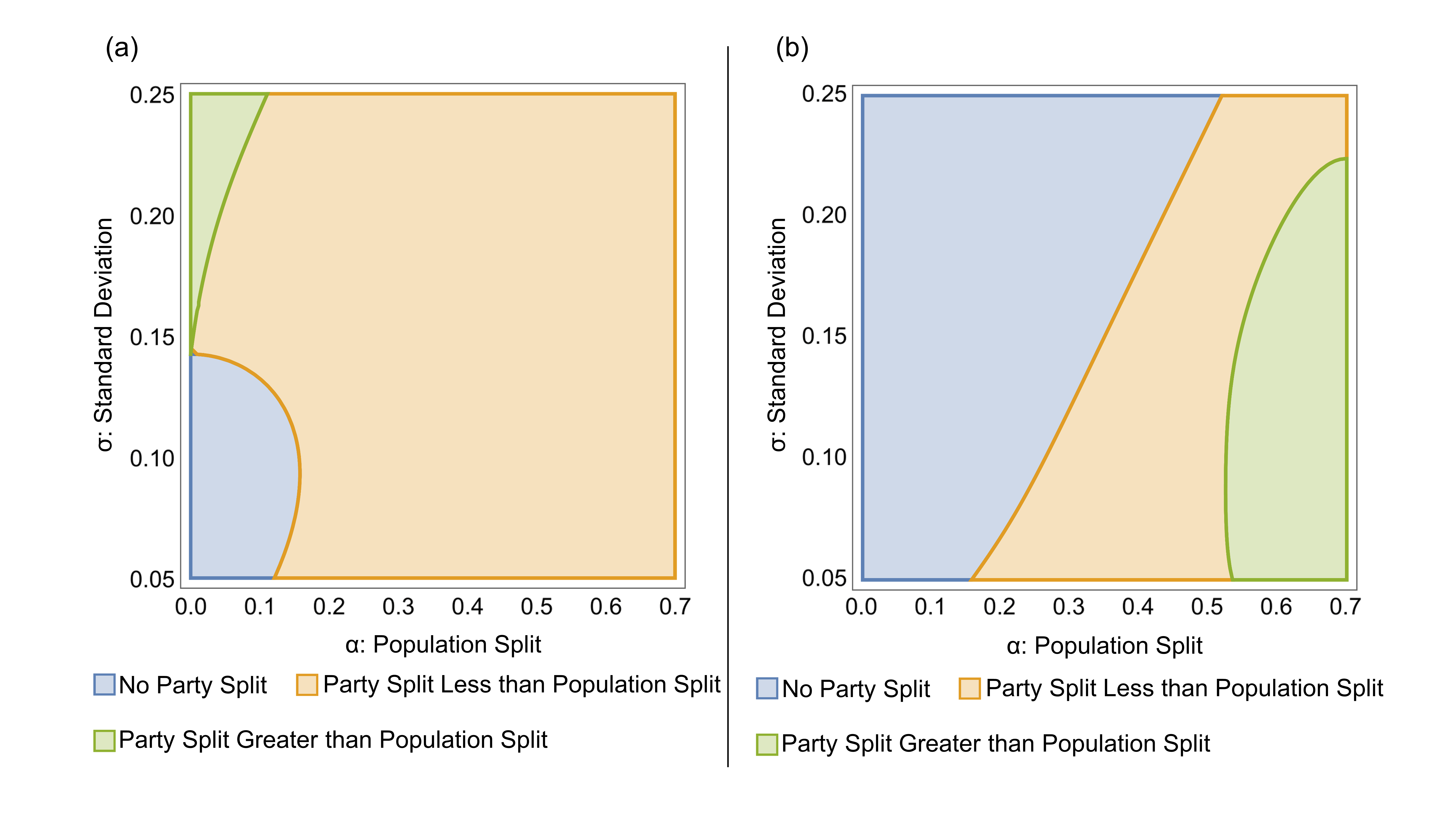}
    \caption{For two sets of model parameters, the nature of the equilibrium candidate positions is shown as a function of the distribution of voter ideology. In both plots, the x-axis is population split and the y-axis is the standard deviation of the two sub-populations. (a) uses parameters $P=2$, $Q=30$, $R=1$, and (b) uses $P=5$, $Q=0$, and $R=5$.
    The three regimes of interest are whether candidates converge to the same position (roughly the mean/median voter theorem result), whether they separate to a lesser extent than the population (the space between the two peaks), or whether they separate to a greater extent than the population. Each space is shaded by adherence to one of these three regimes.}
    \label{fig:phasespace}
\end{figure*}

Depending on voter predisposition to extremist third party candidates, or their willingness to simply stay home in the absence of an appealing candidate, the rational positions taken by main candidates will vary. In our model, candidates qualitatively do one of three things. They either (1) converge to the median similar to standard models of voter choice, (2) deviate from the median but still select positions between the two peaks of public opinion, or (3) deviate from the median to a greater extent than the voting base. Two examples of how voter ideology distribution shapes candidate positions is shown in Figure \ref{fig:phasespace}. For each selected set of sample model parameters, each of the three possible candidate outcomes are possible depending on the ideological spread of voters.

While possibility (2) is interesting primarily because of its deviation from the results typically derived by the median voter theorem, possibility (3) reveals a potential long-term mechanism for voter polarization.  While our model assumes that voter preferences are static and the position of strategic candidates are dynamic, other models have considered the possibility that voter positions eventually come to resemble candidate positions \cite{Kollman1992}. If voter behavior and ideological distribution is one that motivates extremism among rational candidates, this may in turn create a larger spread among voters.

\begin{figure*}[t]
    \centering
    \includegraphics[width = \textwidth]{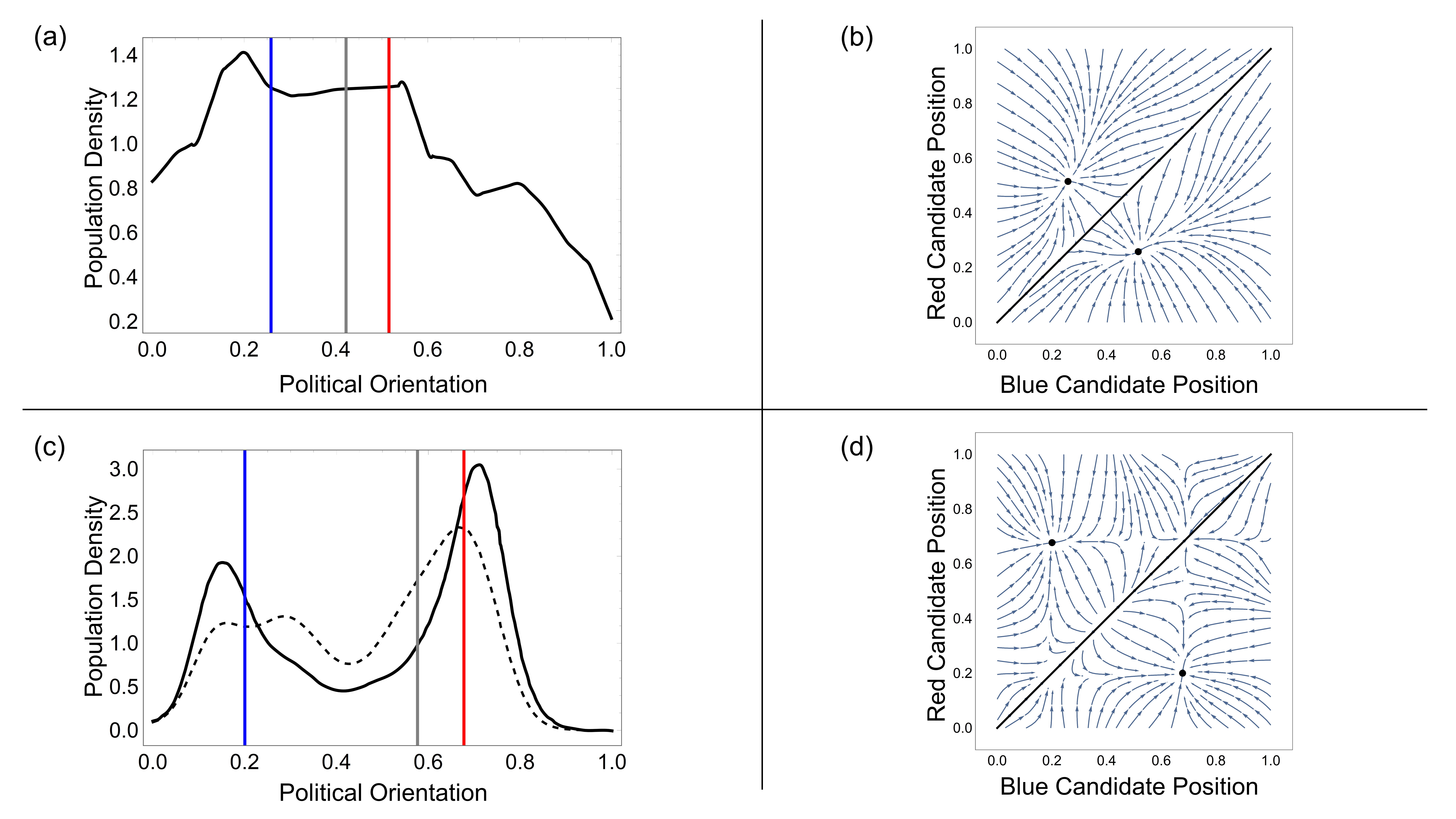}
    \caption{Candidate optimizations based on real-world ideological distributions. First, we show the results of testing our model on real-world data. First, we used the distribution of voter ideology according to the Summer 2017 Political Landscape Survey \cite{pewdataset} (a), and the rational candidate responses to this landscape (b). We also used an analysis of Twitter users \cite{twitterdata} (c), and the rational candidate responses to this landscape (d). In (a) and (c), vertical lines show the convergent position of each candidate (blue and red) and the position of the median voter (grey). (c) also shows the political ideology of political leaders on twitter in dashed grey, and we see that the vertical blue and red lines match nicely with peaks for this curve. Both models used $P=2$, $Q=30$, and $R=1$, a set of parameters that approximate the credence that an average voter may give to not voting or voting for a third party. The population in (c) can be fit to a bimodal distribution. The best fit has the left and right subpopulations' peaks at 0.18 and 0.70, standard deviations at 0.09 and 0.07, and relative weights at 1 and 1.32, respectively. In (a), the population is less bimodal, and so the fit has less value. However, for completeness, we give the values here: positions 0.18 and 0.37, standard deviations 0.04 and 0.41, and relative weights 1 and 81.}
    \label{fig:RealDataPlots}
\end{figure*}

Ultimately, the results of our model show that under very basic assumptions of voters being attracted to third party candidates or prone to staying home, it may make sense for candidates to avoid the center depending on the distribution of voter ideology. We can incorporate observed empirical distributions of voter opinions into a set of model parameters ($P = 2$, $Q = 30$, and $R = 1$) to examine how this model of voter choice might function under contemporary ideological distributions in the United States. Our empirical voter distributions come from two sources.  In Figure \ref{fig:RealDataPlots}a, we see the first data set from \cite{pewdataset}. As we can see, the population here is neither symmetric nor bimodal. However, there is still enough spread in the distribution of the voters to generate a separation between two candidates. The true median of the population ideology is roughly 0.42, but candidates converge to positions at about 0.25 and 0.51. Perhaps unsurprisingly, the asymmetric distribution of voter preferences leads to differing distances between the median position and each of the candidates.

Data from the estimated ideological positions of Twitter users provides a more polarized empirical distribution for examination, and was taken from Figure 3 of \cite{twitterdata} by means of redigitialization. In Figure \ref{fig:RealDataPlots}c we can see that there is a more roughly bimodal distribution, although it remains asymmetric. Once again, political candidates converge on positions that deviate from the median of roughly 0.57, with the left-leaning candidate selecting a position all the way at 0.2, and the right-leaning candidate selecting a position at 0.65. Curiously, while there are more voters on the right than on the left here, the far left positions of the left half of the distribution bring the left-leaning candidate very far from the median. If we suppose that in equilibrium the left choice would win half of the time and the right choice would win half the time, the average position of the winning candidate would be roughly 0.38, very far to the left of the median of 0.57. In this case, the willingness of voters to abstain or vote ``irrationally" for third party candidates gives more weight to the side that entertains more extreme positions.

\section{Discussion}


There is no shortage of proposed mechanisms that explain why voter attitudes have become more polarized over the years. Attitude polarization can result from the twin-mechanisms of homophily, a phenomenon that spans the social and biological sciences \cite{fu2012evolution, mcpherson2001homophily}, and social influence or the diffusion of pairs of associated beliefs \cite{DellaPosta2015,Goldberg2018,DellaPosta2020}. The programming decisions of large media outlets \cite{Campante2013,Prior2013,Sunstein2018}, and the recommendation algorithms of social media sites can send people into wildly different information landscapes \cite{Barbera2020,Levy2021}. The influence these social processes may have on political candidates, however, is less examined. Our model shows that there are very realistic conditions under which rationally behaving major-party candidates will benefit from the amplification of the polarization rather than by strategically pivoting to the center.

This approach, like any model, is limited by the complexity that it emits.  Regarding the specifics of voting in the United States, it omits details on the primary process and how candidates may be bound by verbal commitments they made to a primary electorate while running a general campaign. It omits the possibility of a serious third-party entering the race with strategic rather than ideological motives. It omits the draw that candidate personality may have on the behavior of voters. And it assumes a linear single-dimensional model of ideological positions rather than a multi-dimension~\cite{yang2020us}. It also omits the institutional and geographic complexities of voting induced by district or state-based electoral systems combined with the tendency for voters to self-sort geographically \cite{martin2020sorting}, and strategic attempts to manipulate this process such as gerrymandering \cite{stewart2019information}.

Yet, the minimal number of realistic assumptions necessary to obtain this result makes it all the more compelling and concerning. Stochastically-determined voters with a bimodal ideological distribution and the option to not vote for a major candidate may incentivize more extreme political parties. If we are to believe that voters follow candidates and parties just as candidates and parties follow voters, then a distributional tipping point may exist where voters and candidates chase each other to ideological extremes (see Appendix for an explicit analysis of such tipping point of population split $\alpha$). The solutions to this problem may be found in practices not explored in this model.  For example, ideologically motivated candidates running from the center may effectively ``pull in'' extreme but strategic candidates, in the same way that ideologically motivated extremist candidates can pull strategic candidates away from the center. The polarized political climate in the United States (and elsewhere) remains a serious problem, and continued reconsideration of rational choice voting models with more contemporary assumptions may provide the theoretical material necessary to develop pragmatic solutions for ending what is being referred to by some as a ``cold civil war" \cite{Kay_2021}. 

\section*{Acknowledgements.}
F.F. is supported by the Bill \& Melinda Gates Foundation (award no. OPP1217336), the NIH COBRE Program (grant no. 1P20GM130454), a Neukom CompX Faculty Grant, the Dartmouth Faculty Startup Fund and the Walter \& Constance Burke Research Initiation Award.

\appendix

\section{Closed-form Results for Simplified Model of Ideologically Motivated Voters}
Our model in the main text considers a scenario where individuals decide between voting for a major party, staying home, or voting for a third party. Voters make their choices probabilistically rather than deterministically, with the ideological distance between voters and candidates impacting the weights of behavioral probabilities. This probabilistic decision process together with an abundance of choices make for a main model that is a good approximation of the thought process of the average voter, but is difficult to analyze mathematically. 

While the main model of the voter decision process (with three variables, $P$, $Q$, and $R$) simultaneously captures many plausible elements of voter expression, it makes the calculation of closed form solutions challenging. Here, we examine the specific set of cases when $Q =0$, and $P = R = \ell$ under the limit $\ell \to \infty$. This creates a simplified model of voting where each voter deterministically selects the candidate that is most proximate to them. In this case, everyone votes, and there is no bias in favor of strategic ``major party'' candidates at the expense of ideologically motivated third party candidates. This simplified model would most accurately reflect a population where the costs of voting are effectively zero, and voters are motivated by their ideological similarity to candidates.

Our simplified model presented here lends itself to more tractable functions for the total votes for a candidate, and with a few approximations, allows for closed form solutions. Once again, parties can attempt to maximize votes by making incremental changes to their platform. Now, however, the threat of third parties fixed at both ends of the ideological spectrum are greatly increased, and candidates have a much larger incentive to take polarized positions to motivate their more extreme bases. When the ideological separation between the two centers of political opinion increases beyond a certain point, candidates begin to move away from the median. This is what we refer to as \emph{the first phase shift}. If ideological separation exceeds another point, candidates begin to take on positions that are more extreme than the subpopulations' two ideological centers. This is referred to as \emph{the second phase shift}.

\subsection{Voting Behavior}

Political orientation is represented by a numerical value between 0 and 1. The population's probability density function $f$ has two parameters: a split $\alpha$ that represents the polarization of the population, and $\sigma$, the spread of the population.
\begin{equation}\label{eq:pdfparams}
    f(x) = c\Big( \mathcal{N}(0.5+\alpha/2,\sigma^2) + \mathcal{N}(0.5-\alpha/2,\sigma^2) \Big)
\end{equation}
$\mathcal{N}(a,b^2)$ is the normal distribution with mean $a$ and variance $b^2$.
$c$ is a normalizing constant to ensure that $\int_{0}^1 f(x)dx = 1$. For small values of $\alpha$ and $\sigma$, $c \approx 0.5$ as the normal distributions are close to zero outside the interval $[0,1]$. 
This population is symmetric around $0.5$, and thus the median voter is always located at 0.5.

The blue and red candidates will be located at $b$ and $r$, respectively. We can assume that $b \leq r$. Both parties have perfect information about the population's density function.

\subsubsection{Voter Choice Functions and Candidate Equilibrium}

\begin{figure*}
    \centering
    \includegraphics[width = \textwidth]{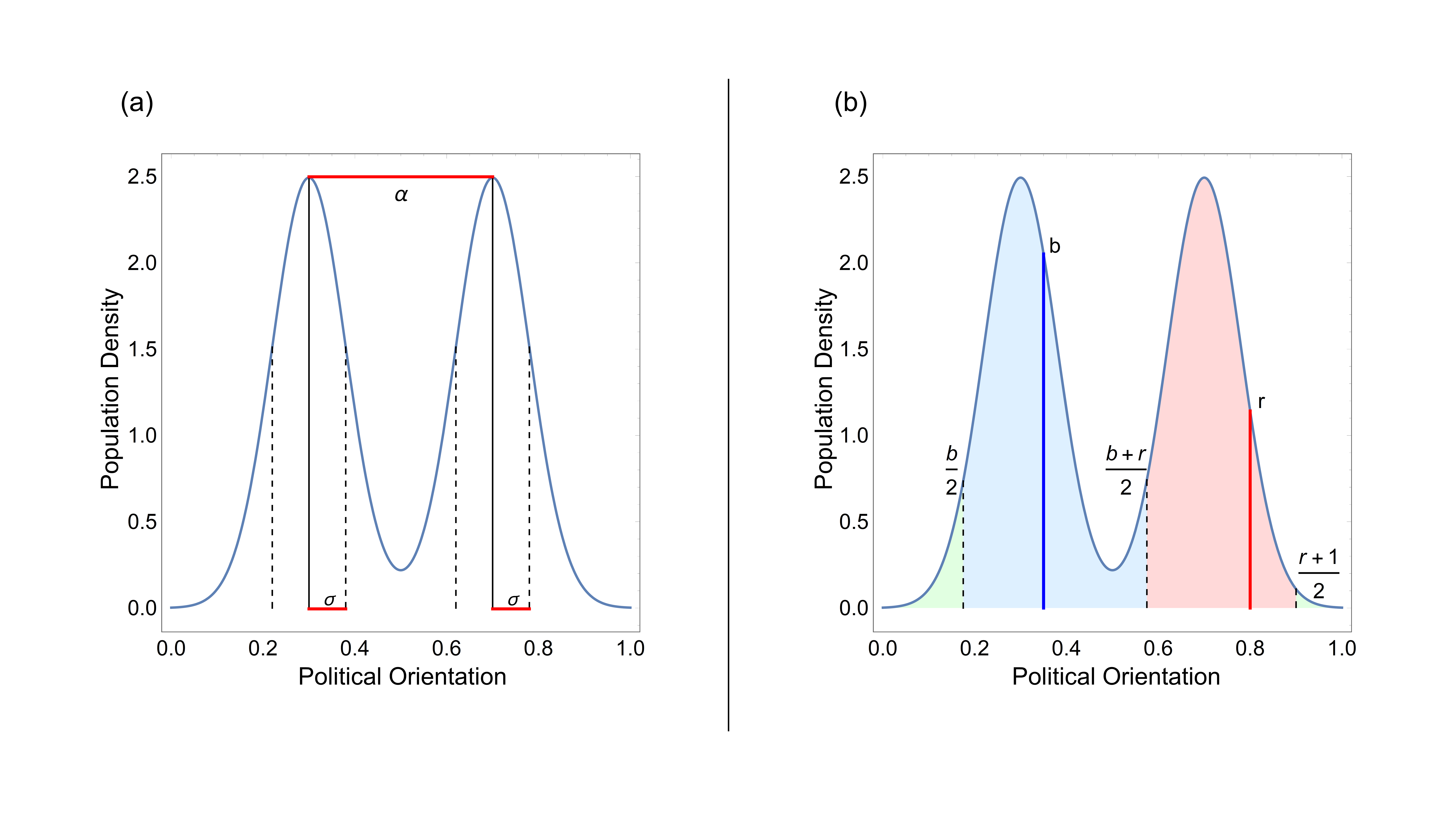}
    \caption{In (a), we see the effects of the two population parameters, $\alpha$ and $\sigma$. $\alpha$ is the distance between the two population centers, and as $\sigma$ increases, the population distributions will become less pronounced and more diffuse. In (b), we see how the population votes when $b=0.35$ and $r = 0.8$. The blue area represents people who voted for the blue candidate, the red area shows people who voted for the red candidate, and the green areas on the edges represent people who voted for a third party.}
    \label{fig:PopParamsVotes}
\end{figure*}

Once the parties are set, the population votes. In the simple model, a voter at $v$ will voter for the closest of $0$, $1$, $b$, or $r$. Therefore, every voter between $\frac{b}{2}$ and $\frac{b+r}{2}$ votes for blue, and every voter between $\frac{b+r}{2}$ and $\frac{r+1}{2}$ votes for red. Up to a constant factor, these quantities can be expressed as integrals of $f$ (see Fig.~\ref{fig:PopParamsVotes}). 

\begin{equation}\label{eq:bluevotes}
    \textrm{blue party candidate votes} = \int_{\frac{b}{2}}^{\frac{b+r}{2}} f(x)dx
\end{equation}
\begin{equation}\label{eq:redvotes}
    \textrm{red party candidate votes} = \int_{\frac{b+r}{2}}^{\frac{r+1}{2}} f(x)dx
\end{equation}

With these simplified equations, we can find a necessary condition for admissible equilibrium. Consider the blue party, whose share of votes is given by \eqref{eq:bluevotes}. The share of votes is dependent on the party platform $b$, so the party can consider adjusting the platform to increase the number of votes. Blue party votes can be maximized by setting the derivative \eqref{eq:bluevotes} to be zero, which leads to:

\begin{equation}\label{eq:blueeq}
    f\Big(\frac{b+r}{2}\Big) = f\Big(\frac{b}{2}\Big)
\end{equation}

When this condition is satisfied, it is possible that the blue party cannot further increase votes by changing their platform slightly. This equation can be used to find the admissible evolutionary stable strategy (ESS) for blue party candidate positioning.

Similarly, the necessary condition for red party votes to  be maximized is
\begin{equation}\label{eq:redeq}
    f(\frac{b+r}{2}) = f(\frac{r+1}{2})
\end{equation}

Therefore,  any given pair $b$ and $r$ that simultaneously ensures an admissible local maximum (ESS) of blue and red votes, respectively, should satisfy the condition:
\begin{equation}\label{eq:eqcondition}
    f(\frac{b}{2}) = f(\frac{b+r}{2}) = f(\frac{r+1}{2})
\end{equation}

\begin{figure*}[!htbp]
    \centering
    \includegraphics[width = \textwidth]{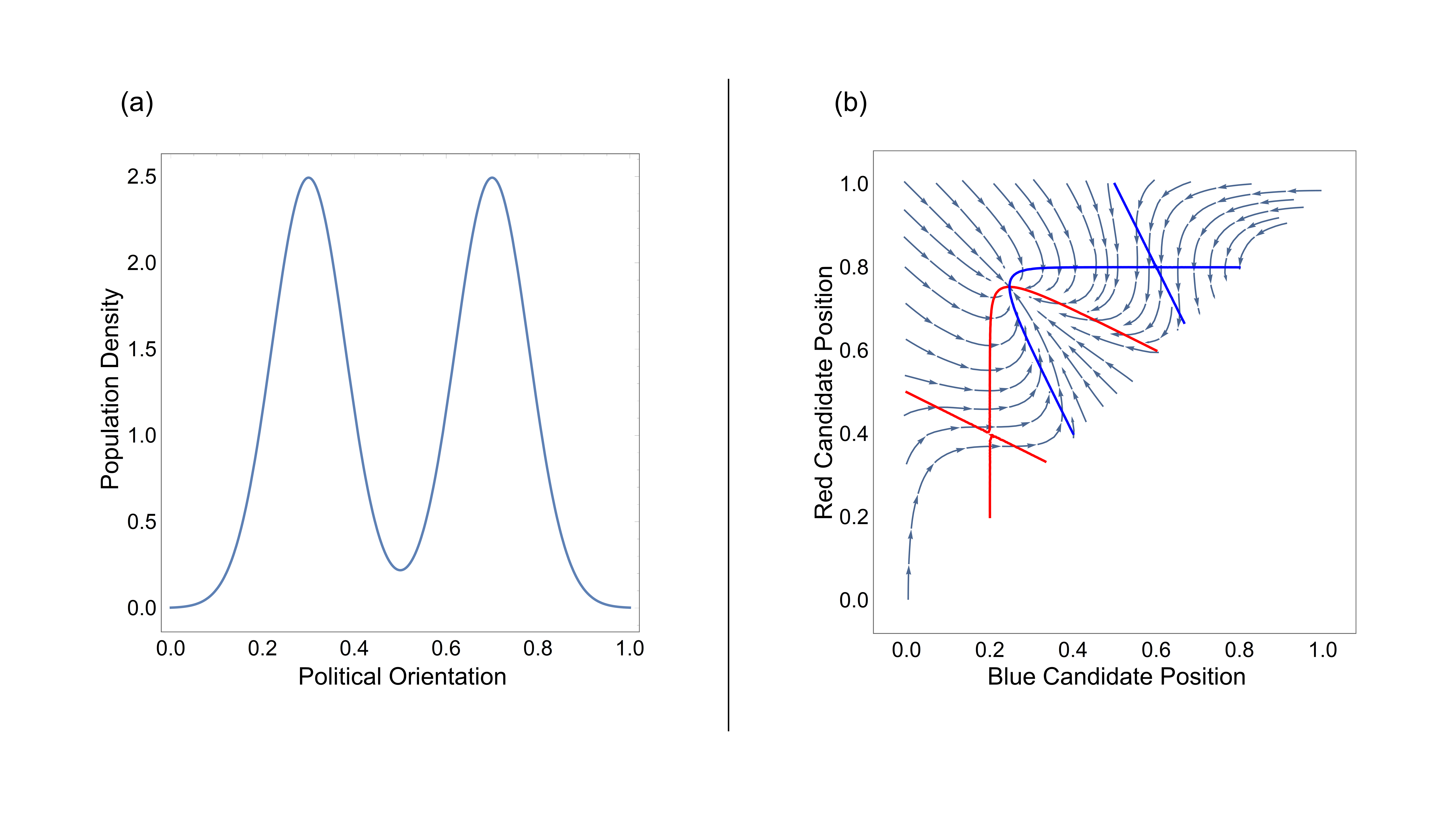}
    \caption{An example of party candidate positioning dynamics within a simple population. (a) shows the density function of a population with split $\alpha = 0.4$ and standard deviation $\sigma = 0.8$. The corresponding candidate positioning dynamics are shown in (b). The blue and red curves are the necessary conditions for admitting evolutionary stable strategies for the blue and red candidates, respectively. The intersection of the blue and red lines show the only \emph{stable} equilibrium point for this population, and the stream plot shows that the parties will naturally move toward this stable equilibrium point from any initial starting pair of ideological coordinates. With loss of generality, we assume $r> b$. }
    \label{fig:JSFdynamics}
\end{figure*}

If \eqref{eq:eqcondition} is not satisfied, at least one party will shift their platform to potentially increase the number of their votes. Figure \ref{fig:JSFdynamics} shows how $b$ and $r$ optimize and counter-optimize through the space of possible ideological positions and ultimately arrive at a local evolutionary stable set of positions.

Multiple equilibrium points can exist, both stable and unstable, depending on the ideological distribution of the population. However, the symmetric, bimodal populations modeled here have a single stable equilibrium satisfying $0.5-\frac{b}{2} = \frac{r+1}{2}-0.5$, that is, we have $ b + r = 1$ (this is because the population distribution $f(x)$ is symmetric with respect to $x = 0.5$). The distance between $b$ and $r$ is dependent on the split and variance of the population. Depending on these parameters, the candidate equilibrium can approach the same position at median voter, have distinct positions that are bounded by the ideological peaks of the population, or have distinct positions that are more extreme than the ideological peaks of the population (see Figures 4 and 5 of main paper). 

\subsection{Analytical Results for Phase Changes}

One result of central interest is when it is strategic for candidates to select positions that diverge from that of the median voter, and furthermore, when it benefits candidates to select ideological positions that are more distant than the two ideological peaks of the proposed bimodal population. We now focus on mathematically identifying the two phase changes between these three possible qualitative outcomes. 

First, observe that the behavior of the candidates is entirely determined by the shape of the population density function $f(x)$, which has two parameters, $\alpha$ and $\sigma$. Here we fix $\sigma$ to be a constant, and consider how changing $\alpha$ affects the equilibrium positions of the candidates. 

When $\alpha =0$, the population is unimodal, and both candidates will unsurprisingly converge on the median, where the density is highest. As $\alpha$ increases, eventually the population density at the median will be surpassed by the population density at $b = 0.25$ and $r = 0.75$, the points halfway between the median and the ideological location of the two third parties (0 and 1).  At this point, appealing to the median voter at the expense of appealing to extreme voters is no longer optimal. Substituting to Equation~\eqref{eq:eqcondition}, this condition can be written as 
 
 \begin{equation}\label{eq:phase1}
    f(0.5) = f(0.25)
\end{equation}

The second phase change will occur at the point where the population centers are equal to the optimal positions for the two opposing candidates, when $b = 0.5 - \alpha/2$ and $r = 0.5+\alpha/2$. To determine when this scenario is an equilibrium, we substitute this into Equation \eqref{eq:eqcondition}, and observe that the second phase change occurs when 

\begin{equation}\label{eq:phase2}
    f(0.5) = f(\frac{0.5-\frac{\alpha}{2}}{2})
\end{equation}

\begin{figure*}
    \centering
    \includegraphics[width=0.8\textwidth]{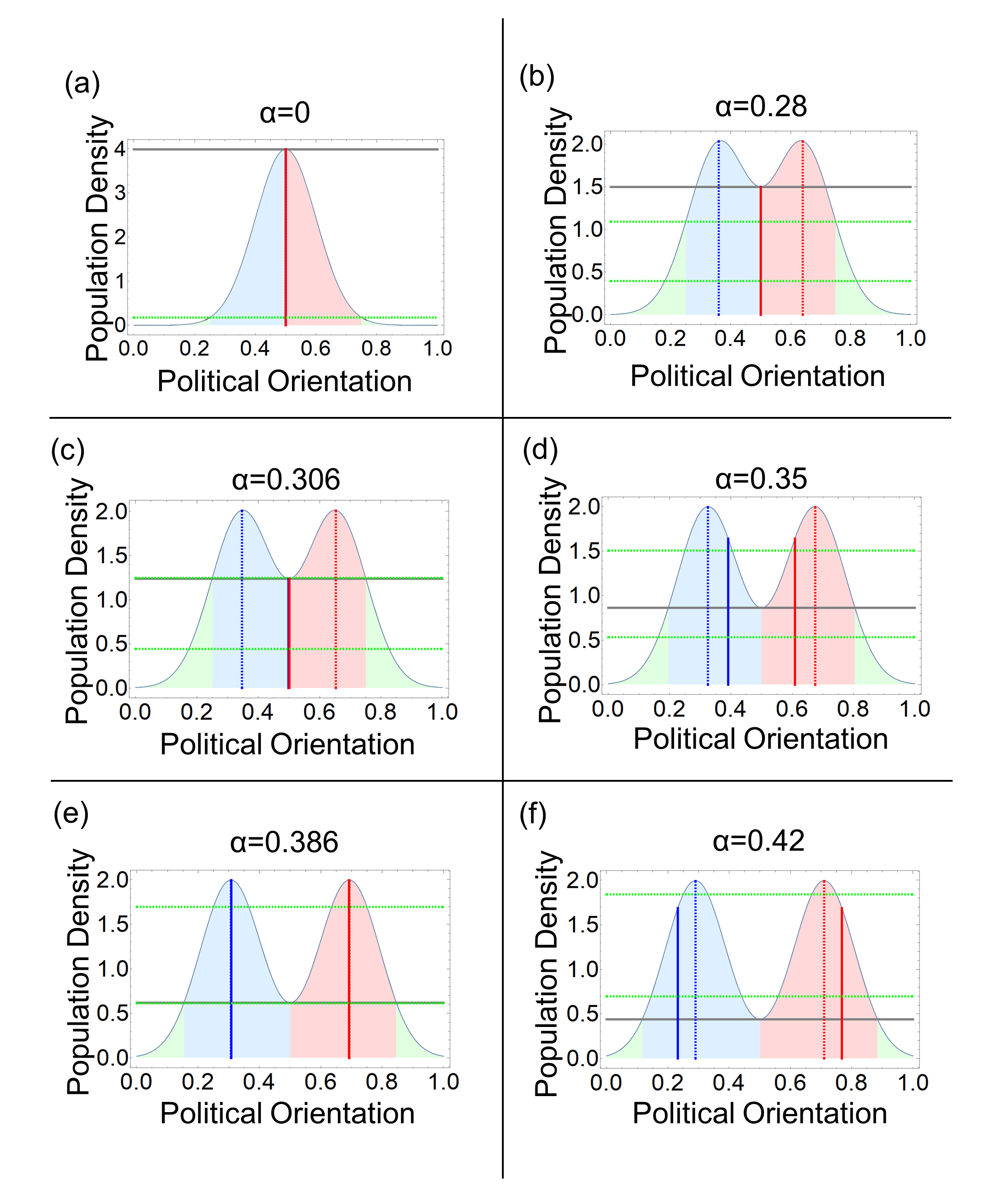}
    \caption{Optimal positions of candidates change as the population becomes more ideologically polarizes (as $\alpha$ increases). The red area reflects vote share for the red candidate, the blue area reflects vote share for the blue candidate, and the green area reflects votes for either third party. The solid blue and red vertical lines indicate the positions taken by the political parties, and the dashed red and blue lines indicate the ideological peaks of the voting population.
    In (a), $\alpha=0$ and the population is unimodal. As $\alpha$ increases, $f(0.5)$ begins to decrease, but is still larger than $f(0.25)=f(0.75)$ and candidates still compete for the median voter. This is seen in (b). In (c) the first phase change is shown: $\alpha$ has grown so that a candidate at 0.5 can take an infinitesimally small step away from the median and not lose any votes. The loss of the median voter is exactly offset by the votes gained at 0.25 or 0.75. In (d), the population density at 0.5 is now less than the density at 0.25 or 0.75. It is now beneficial for candidates to leave the low-density center and appeal to the higher-density areas at 0.25 and 0.75. In (e), the second phase change has been reached, and the optimal candidate position is the same as the population centers. In (f), the population spread is wide enough that the candidates' optimal positions are outside the ideological modes of the population. Model parameters: $\sigma = 0.1$.}
    \label{fig:sketch}
\end{figure*}

Figure~\ref{fig:sketch} illustrates these changes more generally for a fixed level of $\sigma$. In the remainder of this section, we derive the phase changes as a function of both $\alpha$ and $\sigma$.

\subsection{Threshold of Population Split $\alpha$ for First Phase Change}

We now turn to identifying phase changes with regards to both $\alpha$ and $\sigma$. In order to solve Equation \eqref{eq:phase1} for $\alpha$ in terms of fixed $\sigma$, we rewrite Equation~\eqref{eq:pdfparams} as
\begin{equation}
f(x) = c[g(x-0.5-\frac{\alpha}{2}) + g(x-0.5+\frac{\alpha}{2})]
\end{equation}
where $g(x) = \frac{1}{\sigma \sqrt{2 \pi}} \exp{(-\frac{x^2}{2 \sigma^2})}$ is the standard normal distribution with variance $\sigma^2$. 

At $x=0.5$, we have
\begin{equation}
f(0.5) = c \frac{2}{\sigma \sqrt{2 \pi}} \exp{(-\frac{(\frac{\alpha}{2})^2}{2 \sigma^2})}
\end{equation}

At $x=0.25$, we have

\begin{equation}
f(0.25) = \frac{c}{\sigma \sqrt{2 \pi}}  [\exp{(-\frac{(-0.25-\frac{\alpha}{2})^2}{2 \sigma^2})} + 
 \exp{(-\frac{(-0.25+\frac{\alpha}{2})^2}{2 \sigma^2})}]
\end{equation}

Therefore, the threshold of population split $\alpha$ for the first phase change satisfies

\begin{equation}
   2 \exp{(-\frac{(\frac{\alpha}{2})^2}{2 \sigma^2})} = \exp{(-\frac{(0.25+\frac{\alpha}{2})^2}{2 \sigma^2})} + 
\exp{(-\frac{(-0.25+\frac{\alpha}{2})^2}{2 \sigma^2})}
\end{equation}

The above equation can be further simplified to be:
\begin{equation}
e^{-\frac{1+4\alpha}{32\sigma^2}} + e^{-\frac{1- 4\alpha}{32\sigma^2}} = 2.
\end{equation}

$\alpha$ can be solved in closed-form (there exist two solutions of $\alpha$; one positive and the other negative), but is too tedious to be included here. The threshold value of $\alpha$ is given by the positive solution $\alpha_1$.

An approximation of $\alpha_1$ can be obtained if the term $e^{-\frac{1+4\alpha}{32\sigma^2}} \ll 1$, and further by solving $e^{-\frac{1- 4\alpha}{32\sigma^2}} \approx 2$, we get 
\begin{equation}\label{eq:alpha1}
    \alpha_1 \approx \frac{1}{4}+8\sigma^2 \ln{2}
\end{equation}

\subsection{Threshold of Population Split $\alpha$ for Second Phase Change}

We proceed in a similar fashion to identify the threshold $\alpha$ for the second phase shift, by solving \eqref{eq:phase2} for $\alpha$ in terms of fixed $\sigma$.

By equating $f(0.5) = f(0.25-\frac{\alpha}{4})$, we obtain

\begin{equation}
   2 \exp{(-\frac{(\frac{\alpha}{2})^2}{2 \sigma^2})} = \exp{(-\frac{(0.25+\frac{3\alpha}{4})^2}{2 \sigma^2})} + 
\exp{(-\frac{(-0.25+\frac{\alpha}{4})^2}{2 \sigma^2})}
\end{equation}

$\alpha_2$ can be found by numerically solving the equation above. Similarly, an approximation of $\alpha_2$ can be obtained if the term $\exp{(-\frac{(0.25+\frac{3\alpha}{4})^2}{2 \sigma^2})} \ll 1$, and by solving $ 2 \exp{(-\frac{(\frac{\alpha}{2})^2}{2 \sigma^2})} =
\exp{(-\frac{(-0.25+\frac{\alpha}{4})^2}{2 \sigma^2})}$, we get

\begin{equation}\label{eq:alpha2}
    \alpha_2 \approx -\frac{1}{3} + \frac{2}{3}\sqrt{1+24\sigma^2\ln{2}}
\end{equation}

\subsection{Population Parameters and Candidate Behavior}

\begin{figure*}
    \centering
    \includegraphics[width = \textwidth]{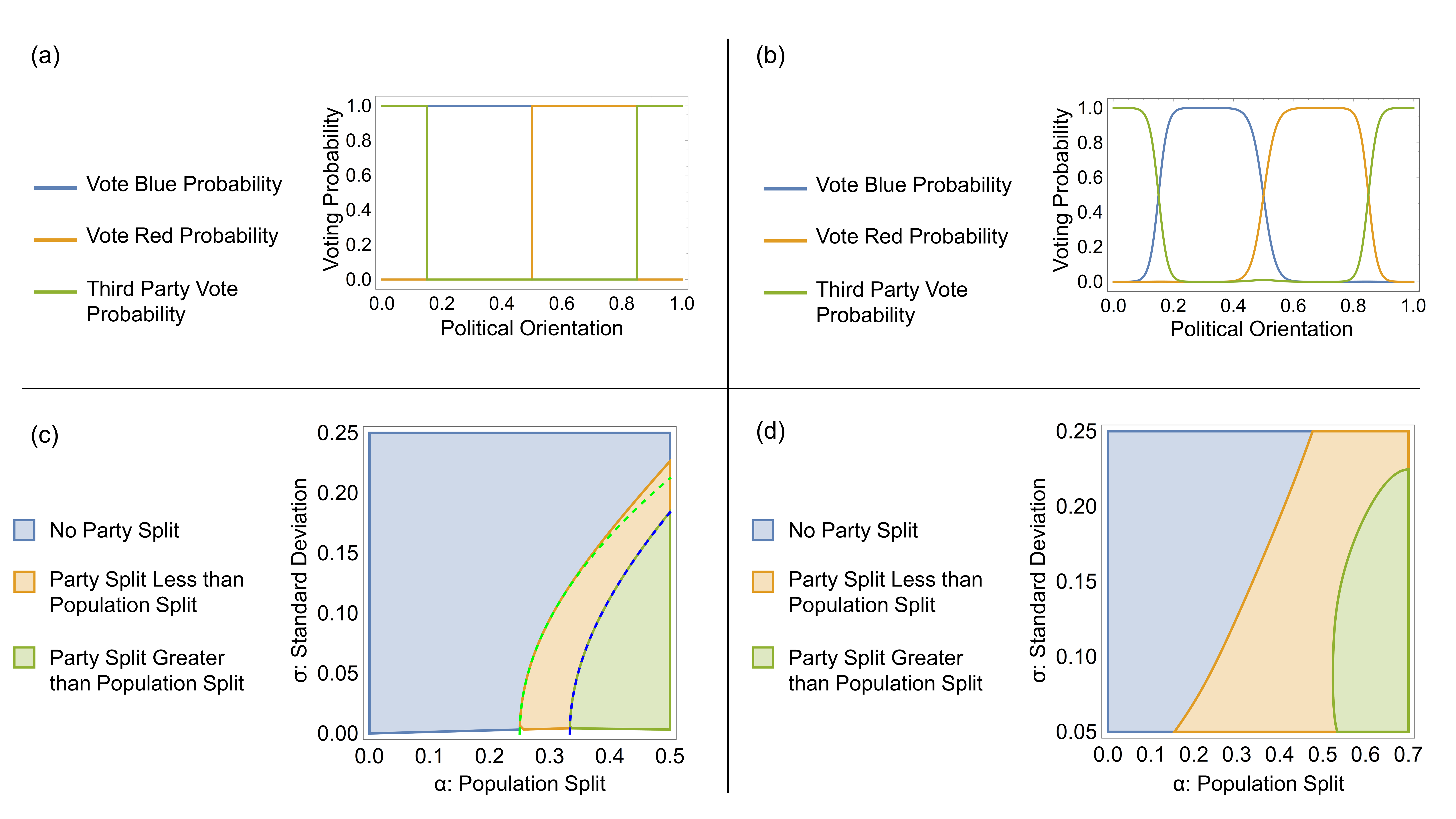}
    \caption{Voting probabilities and corresponding regions of behavior for the simplified model presented here and the main model with parameters $P=R=5$ and $Q=0$. In (a) and (b), we see how $P=R=5$ and $Q=0$ gives a rough approximation of the simplified model's voting behaviors with candidates at 0.3 and 0.7. In (c), we see how population structure affects the equilibrium position. The dashed curves show the analytic approximations of the phase changes (equations \eqref{eq:alpha1} and \eqref{eq:alpha2}), which match the actual phase changes very well. As the standard deviation increases, the sub-populations become more diffuse and the approximation becomes less accurate. (d) shows similar behavior as (c), with differences between the two being explained by the roughness of approximation demonstrated in (a) and (b).}
    \label{fig:regionplots}
\end{figure*}

Now that we have full equations for both of our phase changes, we can examine how the shape of the population density function affects the location of the candidate position equilibrium. Here we compare the deterministic model presented in this section where voters choose the most proximate candidate, and the model presented in the main text where $P=R=5$ and $Q=0$. In this case voters never stay home, and voters given third-parties the same consideration as major parties, but they do not always choose the most ideologically proximate candidate. We visually compare the results between these two models in Fig.~\ref{fig:regionplots}.

In Fig.~\ref{fig:regionplots}c, we see that for a fixed standard deviation, as the population split increases, the density at the median goes down and the candidate split increases from 0 to greater than the population split. As the standard deviation for each of the two underlying distributions ($\sigma$) increases, the population becomes more diffuse, and the density around the median voter remains large, encouraging candidates to compete for the middle and allowing more extreme voters to choose an extreme third-party. While the two region plots have quantitative differences, they are qualitatively similar. In both Fig.~\ref{fig:regionplots}c and Fig.~\ref{fig:regionplots}d there is a narrow, diagonal band in which the party split is non-zero but less than the population split. Interestingly, this comparison also suggests that stochastic voter decision making widens the range of population ideology distributions that lead to outcomes in the ``middle" phase space. That is, voter stochasticity may incentivize candidates to adopt differing positions, but not positions that are more extreme than the bimodal centers of the electorate.

%

\end{document}